# Agile Information System Development Organizations Transforming to Large-Scale Collaboration

*Completed Research Paper*

**Marius Mikalsen**
SINTEF and Norwegian University of
Science and Technology
Trondheim, Norway
marius.mikalsen@sintef.no

**Nils Brede Moe**
SINTEF
Trondheim, Norway
nils.b.moe@sintef.no

**Sut I Wong**
BI Norwegian Business School
Oslo, Norway
sut.i.wong@bi.no

**Viktoria Stray**
SINTEF and University of Oslo
Oslo, Norway
stray@ifi.uio.no

## Abstract

*We report findings from a case study of a large agile information systems development (ISD) organization´s sudden transformation to distributed, digital work in the context of the Covid-19 pandemic. It seeks to understand how knowledge creation and sharing changes. The findings show various forms of distance being introduced, digital tool usage, increased task orientation, and variations across teams. To analyze the findings, we use the concepts of large-scale collaborations and sociability. Large-scale collaboration offers a socio-technical perspective on tackling distributed knowledge sharing and creation in the presence of multiple, loosely coupled partners using digital tools for collaboration. We show what the digital tools afford using the concept of sociability. We discuss how distributed digital practices make teams more task-oriented and that creating and maintaining sociability, a key issue for knowledge sharing in agile ISD organizations, require relation-oriented communication during practical problem solving using digital tools.*

# Introduction

The Covid-19 pandemic has caused organizations to transform from co-located to distributed work using digital tools. For agile Information System Development (ISD) organizations, which typically relies on co-located teams, remote working introduces challenges for knowledge creation and sharing. In particular, knowledge creation and sharing is largely tactic and intimately tied to the ways in which agile team members discuss and solve problems together (Chau et al. 2003; Ghobadi and Mathiassen 2016). Social ties and common ground are crucial to their knowledge creation and sharing (Kotlarsky and Oshri 2005). Against the importance of physical proximity to establish team members´ social ties and common ground, we investigate the role of digital tools when a large agile ISD organization is suddenly being forced to be distributed due to a pandemic and must collaborate, create, and share knowledge exclusively using digital tools. This is pertinent given known knowledge challenges with distributed agile ISD (Sarker and Sarker 2009).

We therefore ask the research question: How do digital tools afford knowledge creation and sharing in agile ISD organizations in a sudden transition to distributed work? To answer this question, we report findings from a case study of a large, agile ISD company with 24 teams that provides digital solutions to banks. The teams in the company were primarily co-located and, literally overnight, had to transition into completely distributed, digitally mediated work. Our results show how their practices changed, details on their tool usage, the differences between teams, how the lockdown influenced the relationships with teammates, and crucially, how the digital tools played a role in this process.

To analyze our results, we take as our starting point the concept of large-scale collaborations, or, LSCs (Malhotra et al. 2021). LSCs are larger pools of loosely coupled participants (from different organizations, or within the same organizations) using a common digital platform for knowledge creation and sharing in carrying out a common mission. Our motivation for using the concept is that LSCs are used to address large and complex problems such as open-source operating systems. LSCs often face challenges in reaching a shared understanding and common agreement, i.e., knowledge deliberation. To overcome these challenges, it is necessary to consider the socio-technical affordances of the digital tools in LSCs. This implies to understand what digital tools afford in terms of allowing LSCs to deal with the variances that hits them. LSC research does not provide much consideration of social ties and common ground, however. To add to this, as agile ISD organization by practical necessity are likely to want to maintain social ties and common ground despite becoming distributed, we apply the concept of sociability. Sociability captures the extent to which digital tools are perceived to facilitate the emergence of a sound social space with attributes such as trust, belonging, a strong sense of community and good working relationships (Kreijns et al. 2007). The sociability framework thus focuses on the degree to which digital tools afford relational-oriented communication while also achieving task-oriented communication. We contribute by reporting empirical material about an ISD organization's response to the sudden change into distributed, digital practices. Also, using LSC and sociability as a lens, we discuss how the digital tools used affords agile ISD organizations to become more task oriented and formal, how different teams have different variations that needs to be addressed, and how agile ISD organizations should work to maintain and create sociability using digital tools for knowledge creation and sharing.

The rest of this article is organized as follows. Section 2 explains the transition from traditional ways of considering knowledge creation and sharing in agile organizations, to what a LSC and sociability perspective entails. Section 3 gives the case background and methods used in the study. Section 4 provides the findings, including changing practices, tool usage and exploring the teams' sociability scores. Section 5 discusses these findings and infers that becoming completely digital marks a formal turn at the expense of more social aspects of teamwork and that it is key to find ways to maintain and create new sociability through digital tools. We also discuss the practical implications of our findings. Section 6 concludes the paper.

# Theory

## *Knowledge Creation and Sharing in Agile ISD Organizations*

ISD organizations seek to react to the market, respond and incorporate changing customer requirements, and, simultaneously deal with technological innovations, and the concept of agility is increasingly being seen as relevant (Börjesson and Mathiassen 2005). Some even call it the root metaphor for an organizational strategy towards digital innovation (Berente 2020). The information systems (IS) discipline´s understanding of agility concerns an organization´s capability to sense, respond, learn, and in that manner, improve the delivery of value. The core of agile ISD organizations ability to anticipate, create, learn from and respond to user requirements changes – has been described as 4 concepts: (1) incremental design and iterative development, (2) inspect and adapt cycles, (3) continuous customer involvement and (4) working cooperatively/collaboratively/in close communication (Baham and Hirschheim 2021). Similarly, Qumer and Henderson-Sellers (2008) define agility in the context of ISD as a method that is people-focused, communication-oriented, flexibly, speedy, lean, responsive, and learning.

Variability and continuous change are at the core of such depictions, and knowledge creation and sharing are intimately tied to change. Discovery and adoption of digital products happen through agile sense and respond capabilities (Lyytinen and Rose 2006). Agile organizations consequently seek to embrace variation by learning from change in order to contribute to customer value (Conboy 2009). Knowledge creation and sharing is tied to practice, what Orlikowski (2002) calls knowing in practice. Here, knowing is an ongoing social accomplishment, constituted and reconstituted in the everyday work practices of dealing with change and variation. With its practice focus, knowledge creation and sharing in agile organizations is situated and emergent. There is a focus on minimum documentation, requirements are understood by close contact with the customer, training is done by pair programming, competence management is done by daily meetings, and continuous learning is facilitated through retrospectives (Chau et al. 2003). Team factors (such as diversity), process factors (such as communication and organization), and contextual factors (such as project setting and technology used), may all be barriers to knowledge creation and sharing (Ghobadi and Mathiassen 2016). Knowledge sharing is tied to collaboration and mutual adjustment through frequent synchronisation meetings (e.g., daily stand-ups). The theoretical underpinnings of agile methods focus on regular reflection for continuous knowledge creation and sharing (Babb et al. 2014). Reflection happens *in*-action, such as release planning, group programming, and demos. Reflection also happens *on*-action, such as through daily standups and retrospectives.

### *The Challenges of Knowledge Creation and Sharing Over a Distance*

Knowledge creation and sharing based on common ground and social ties are challenged when different forms of distance are introduced. Research on distributed ISD has shown the importance of the relationships between social ties, knowledge sharing and successful collaboration in distributed teams (Kotlarsky and Oshri 2005). Relationships and networking are essential when solving complex, unfamiliar, or interdependent ISD tasks. Network size and networking behavior depend on company experience, employee turnover, team culture, need for networking, and organizational support (Šmite et al. 2017). Aspects of social capital, such as shared understanding and relational trust are crucial for successful knowledge sharing in distributed ISD contexts (Zimmermann et al. 2018).

Knowledge sharing and common ground is even more important in agile distributed ISD as it relies on mutual adjustment and frequent communication. For knowledge sharing to succeed, effective communication is a key success factor to remedy that of not being co-located and seeing and experiencing what others are doing (Shameem et al. 2018). Effective knowledge sharing implies the abilities to be aware of the actions of distributed team members, utilize cultural differences and maintain connectivity with customers (Sarker and Sarker 2009). However, cultivating effective communication is, not surprisingly, more challenging due to physical, temporal, socio-cultural, knowledge and experiential distances (Ghani et al. 2019). The use of digital communication tools due to distance also brings challenges, such as reduced communication quality as a result of poor network connections and meaning, tone and emotion being lost and/or misunderstood over digital media with fewer capacities to convey highly contextual communication (Rizvi et al. 2015).

Despite having significant tool support, such as knowledge repositories and databases, the role of social knowledge networks is significant in order to look for actionable knowledge to solve the problems at hand (Balijepally and Nerur 2019). In distributed ISD settings research finds that dedicated knowledge sharing tools, such as wiki pages are not used or are found unsuitable. Tools that facilitate knowledge sharing are tools that support collaborative practices, such as tools for pair programming, teleconferencing and online chats (Anwar et al. 2019). While a number of mechanisms aim to support knowledge creation and sharing through social networks, for example, facilitation of communities of practice (Smite et al. 2019) and adequate communication tools (Stray and Moe 2020), organizations still experience challenges. In large-scale distributed agile ISD contexts, even mature agile teams experience the usual barriers such as language, unbalanced activity levels, and difficulties balancing the communication when using new tools. One challenge is the use of team members' direct messages when using tools such as Slack, because direct messaging increases the response rate but reduces transparency of the communication to others. Further, while Slack channels, Google groups, Trello boards, and other tools provide valuable means for interaction, face-to-face sessions were found to be more important for boosting knowledge sharing (Smite et al. 2019).

### Towards a Large-Scale Collaboration Perspective on Knowledge Creation and Sharing in Distributed Agile Organization

Against this backdrop of knowledge creation and sharing challenges with distributed ISD and distributed agile ISD, it is necessary to investigate how it is at all possible to achieve. Pushing beyond the notion of common ground we find that the concept of large-scale collaborations (LSCs) may help explaining what happens to distributed and digitally mediated knowledge creation and sharing. LSCs research address the socio-technical aspects of distributed collaborations (Malhotra et al. 2021). LSCs are larger pools of participants using a common digital platform for knowledge creation and sharing in carrying out a common mission. LSCs are used to address large and complex problems such as scientific discovery within physics (Tuertscher et al. 2014) and open source software (Lee and Cole 2003; O'mahony and Ferraro 2007). LSC practices can also reside within a company (Stol and Fitzgerald 2014). The reason LSCs are possible is because of the Internet and an ever-increasing availability of new digital tools supporting loosely coupled, digital collaboration at scale. An LSC perspective is relevant to understand agile organizations transforming to distributed work, because research that applies small group assumptions focused on social ties and common ground may not properly explain what can be achieved in LSCs (Faraj et al. 2016; Rolland et al. 2016).

Knowledge deliberation, i.e., to be able to discuss shared knowledge in order to increase understanding of a problem and to stimulate solutions, is different in LSCs as they typically are not based on social ties and common ground. Deliberations in LSCs are not resolutions, but rather discussion around a topic intended to increase a shared understanding about the topic, and then form a common agreement about the topic (Malhotra et al. 2021). Productive deliberations include reflective inquiry, abstract reasoning, debate, and dialogue on topics essential to the work being done. The important aspect for LSCs is not to eliminate the need for deliberations, but to design practices and tools that support deliberations. Consider the example from a study of Lee and Cole (2003) of the Linux kernel development. 77% of software patches submitted were not accepted by the core group, and if such rehects occurred without deliberation, participant considered the rejections baseless. Similarly, Stol and Fitzgerald (2014) finds that differences of opinion might negatively affect a project as a large number of stakeholders might lead to conflicting requirements and personal conflict. Deliberations are typically harder in LSCs because of the lack of social ties and common ground. It can be difficult to to know if conflict comes from productive task conflict, or interpersonal conflict, participants are not only representing their own interest, but also that of their team or organization, and there is a lack of shared memory because of participant throughput (Malhotra et al. 2021).

To overcome deliberation challenges, it is necessary to consider the socio-technical affordances of the digital tools in LSCs (ibid.). This implies a focus on technology and to investigate what it affords. In the context of deliberations in LSCs, an affordance can be conceived as a relation between LSC participants´ need to accomplish knowledge deliberation, and the features of collaboration technology used by the participants. Socio-technical systems theory posits that variances (i.e., deviations from expectations or goals) in inputs, processes, and outputs exist in all work systems (Trist and Bamforth 1951). These variances relate to knowledge deliberation in LSCs. Increasing the variation can nurture diversity and divergence, thereby stimulating new ideas and innovation. Reducing variance standardizes processes, hence increasing speed and efficiency. Whether variances should be increased or reduced depends on the variation required in the organization (exploitation versus exploration for example).

As described above, agile organizations traditionally manage variances by practicing knowledge deliberation through being collocated, building social ties, and creating common ground. So, when an agile organization transitions towards LSC, it is reasonable to believe that it wants its digital tools to afford not only task related deliberations (which are most found in LSC research), but also to afford maintaining the social ties and common ground relevant to the knowledge deliberations (Balijepally and Nerur 2019; Kotlarsky and Oshri 2005). The capability of digital tools to afford this is perhaps best illustrated by online community research where they have found how online communities, because of their reliance on the "narrower" digital means of communication, traditionally have been compared to the gold standard of traditional face-to-face interactions (Faraj et al. 2016). This leads to observations as we have described above from distributed agile ISD, where virtual communication fails to mimic face-to-face communication. Still, it is suggested that it can be more rewarding to understand the "surprising width and depth of sociality unfolding" in online communities than to focus on what they fail to accomplish (ibid.).

Agile ISD organizations, when suddenly becoming distributed (as was the case during the Covid-19 shutdown), take on a form of LSC. Then it becomes necessary to consider the role of the digital tools used, and what they afford in handling variations. Traditionally, the focus in LSCs has been task focused, but recent research from distributed agile ISD and online communities reminds us that social ties and common ground can play a pivotal role in distributed and digital knowledge creation and sharing. Faraj et al. (2015) find that, among others, sociability (in terms of social sign off, such as saying goodbye, saying thanks, sharing stories, and referring to individuals by name) lead to being identified as a leader by members of online communities. Sociability in technology "represents the degree to which an individual's desire to socialize is satisfied through a system that is able to provide social interactions with others" (Junglas et al. 2013). Sociability has been found to apply both to knowledge creation and sharing in online communities (Phang et al. 2009). It is therefore necessary to investigate the transition of an agile ISD organization towards a LSC regarding how well digital tools supports both task-oriented and social-oriented communication. To do that, we use the concept of sociability, which refers to the extent to which digital tools are perceived to facilitate the emergence of a sound social space with attributes such as trust, belonging, a strong sense of community and good working relationships (Kreijns et al. 2007). The framework thus focuses on the degree to which digital tools afford relational-oriented communication but also acknowledges their importance in achieving task-oriented communication.

## Case and Method

We chose a revelatory case study (Yin, 2014) to investigate the research question of what happens to communication and trust when many co-located agile ISD teams suddenly become distributed. We studied a case where work was co-located before the Covid-19 pandemic shutdown. Our case consisted of 24 teams in an ISD company delivering digital solutions to an alliance of Norwegian banks. The company was described by practitioners as a leading ISD environment in Norway, using state-of-the-art methods and technologies and regularly presenting their work at national ISD conferences. In our sample, 207 people went from predominantly on-site work to 100% virtual teams. Four of the 24 teams were partially distributed before, with some team members working in another city. The study is revelatory because, to our knowledge, few empirical studies exist on this topic, and we seek new insights to generate ideas for research agendas addressing sudden shifts into digital work and to inform future forms of hybrid work (mixes of co-located and digital work) for agile ISD teams.

### *Case Context: A Co-located Agile ISD Organization Turning to 100% Digital Work*

The 24 ISD teams in our sample were from different domains, such as finance, core data and machine learning. Each team had personnel with different roles, such as software developers, testers, designers, user experience (UX), team leads and product owners. In recent years, they had worked on moving away from their legacy monolithic technical architecture, which is typical for banks, towards a microservice architecture, allowing continuous development for the teams to have end-to-end responsibility for their products and no handovers between teams.

The teams worked with different agile methods depending on what problems they were addressing. The teams had considerable freedom in how they worked. Most teams used a Kanban variant with Scrum elements and several deliberation practices, such as backlog meetings, team meetings and daily stand-ups. They used objectives and key results (OKRs) to guide their work and, influenced by Wodtke (2017), used "Monday commitments" and "Friday wins". They also regularly performed team health checks where team members reported on aspects of the teamwork, such as the clarity of their goals. Additionally, team leaders usually follow up with team members in one-to-one conversations. They used retrospectives to improve work practices, and they used structured problem solving for continuous improvement.

The teams used a suite of digital collaboration tools to support work within and across teams, including the project-management platform Jira, the software version control services GitHub and Maven, Confluence for knowledge sharing, Trello for task management, the messaging platform Slack, Snapchat groups for social sharing, and Microsoft Teams.

### Data Collection and Analysis

We chose a mixed-method approach for this study, combining both qualitative and quantitative methods. We used three types of data collection: a survey, semi-structured interviews, and documents (see Table 1). First, we distributed the survey in April 2020 through Qualtrics software. From the 311 team members on 24 teams to whom we distributed the survey, we received 226 responses: a 73% response rate. Second, a week after the survey was distributed, we conducted an initial analysis of the responses, comparing the 24 teams to observe their communication patterns. As part of a larger survey, we measured sociability, team coordination, electronic dependence, which have been tested and validated by previous studies, as shown below. Besides they are validated measures, these measures were chosen also because of the conceptual fit to our study. In addition, we asked participant to provide some demographic information. All items except demographic questions were measured using five-point Likert scales. We conducted exploratory factor analyses and regression analyses using the analysis software SPSS. We presented the findings back to the case to get feedback and used interviews and documents to get more in-depth insight into findings from the survey.

We used the following measures:

- Sociability: We adapted Kreijns et al. (2007) sociability metrics, modifying the items to fit our context. We changed "computer-supported collaborative learning" to "digital work situation". The measure was chosen to shed light on how sociability of digital tools was perceived.
- Coordination: We measured coordination with five items from (Lewis 2003) Transactive Memory System Scale. Sample items included "Our team worked together in a well-coordinated fashion" and "We accomplished the tasks smoothly and efficiently." The measure was chosen to consider implications for coordination.
- Electronic dependence: We used (Gibson and Gibbs 2006) scale to capture the degree to which individuals depended on digital tools to stay in touch with their team members in their work. The items are much used for measuring dimensions of virtuality such as electronic dependence. A sample item was "To what extent do you rely on e-mail in your daily work?"

| Data source | Amount | Description |
|---|---|---|
| Survey | 226 respondents | All roles, developers, UX, designers, team leads, development leads, etc. |
| Interviews | 12 interviews | We interviewed people from the case, such as team leads and development leads. |
| Documents and logs | N/A | Slack usage data, internal communication from management (Wiki), documents on cycle times, internal survey (health check), presentations. |

**Table 1. Data Sources and Description**

# Results

### Increased Distance, Formal Meetings and Stable Production

On Friday, March 13, 2020, all schools and kindergartens in Norway closed due to the Covid-19 virus outbreak, and home schooling started. The bank (together with most other Norwegian businesses) decided to close all its offices, and the teams had to work from home and be completely distributed. As schools closed, people with children at home had to attend to them and do home schooling, resulting in fewer working hours or shifting work hours to the evening. An interaction designer explained, "It's like working in another time zone", pointing out a temporal distance with the team. Others explained that it was difficult to find work-life balances and that they often worked overtime, as there was nothing else to do staying at home.

Some employees experienced "flow" because of fewer meetings and interruptions. One team lead reflected positively on the new situation: "The physical distance has [before Covid-19] been a challenge. You are closer to some teams and some stakeholders. Now the distance is the same for everyone", pointing out the equality of the physical proximity enabled in the digital environment. However, one designer experienced the home working situation differently: "The home-office situation radically enhances existing negative tendencies in the team, such as a lack of communication and dynamic".

All teams were encouraged to keep their practices (e.g., meetings and ceremonies) and actively experiment with new practices and adapt to the home office situation. Teams in the company were already using digital tools for collaboration before the Covid-19 lockdown, such as Slack for communication in and between teams and a digital software production toolchain with Maven and GitHub.

Previously unscheduled meetings now needed to be scheduled. Meetings scheduled now started on time instead of five minutes late, unlike before Covid-19: "You just push a button, and then you are in the meeting." On average, the respondents spent 2.3 hours in planned meetings and 1.1 in unplanned meetings per day. Differences existed between teams in time spent in meetings, however. Some teams used 0.5 hours per day on ad hoc meetings, while others used over 1.5 hours and others over 3 hours. One team leader gave some insight into the differences: "The informal meetings are becoming formal – you have to put them into the calendar." This could imply that teams that used to have more informal meetings may have reported more planned meetings.

To investigate how working from home affected software productivity in the company, we looked at cycle time (the number of days a task is in progress), waiting time (waiting for another task to be finished) and days used testing. The data from 2018 until June 2020 show a consistent decline in cycle time (see Figure 1), and there were no observable changes during this sudden lockdown period. The waiting time did not change.

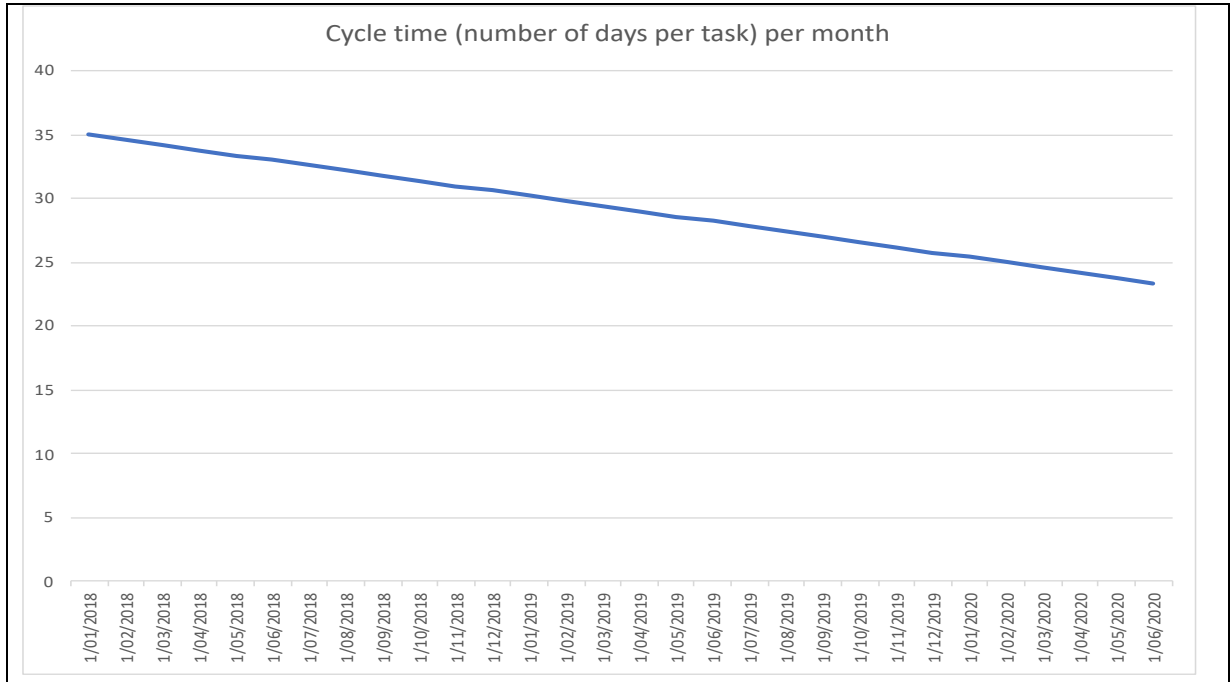

**Figure 1. Cycle Time Indicating Steady Production Levels.**

### *Tool usage: Predominantly private communication*

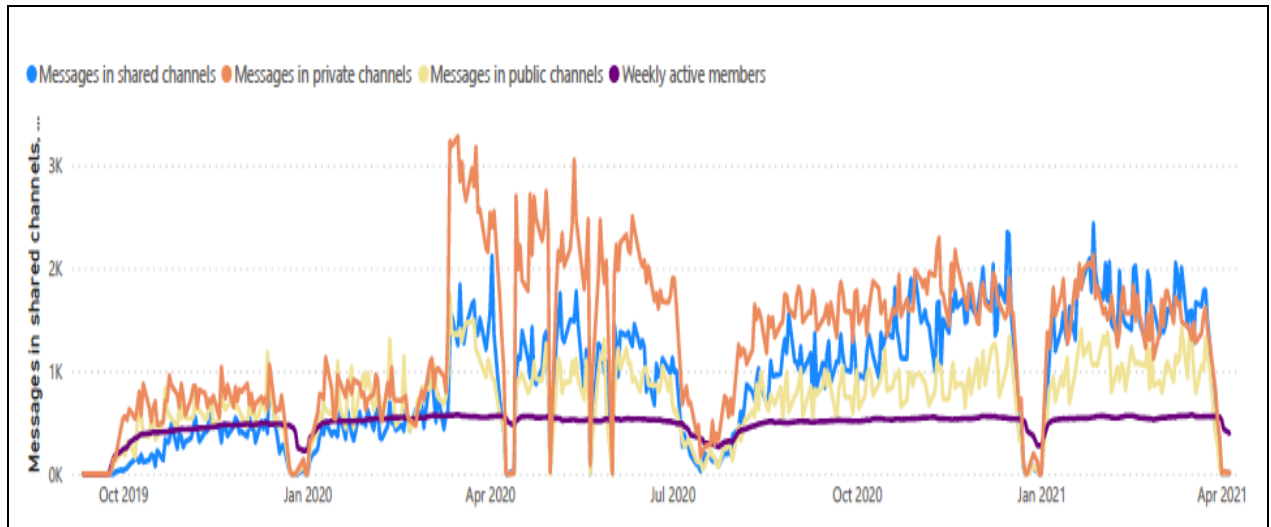

**Figure 2.  Messages in shared private and public channels and weekly active members**

Figure 2 illustrates the changes in Slack usage. The graph shows that on March 12, as the pandemic hits, usage increased dramatically. There are predominantly private channels that are used, that is, channels where users must be invited and those are not searchable for people outside the channel. Such channels are typically used within teams. Public channels are available for everyone within the organization, and users can search for information and join these channels freely. Shared channels are channels between two different organizations. Figure 3 shows direct messages. That is, messages sent directly to another person or to a group of persons.

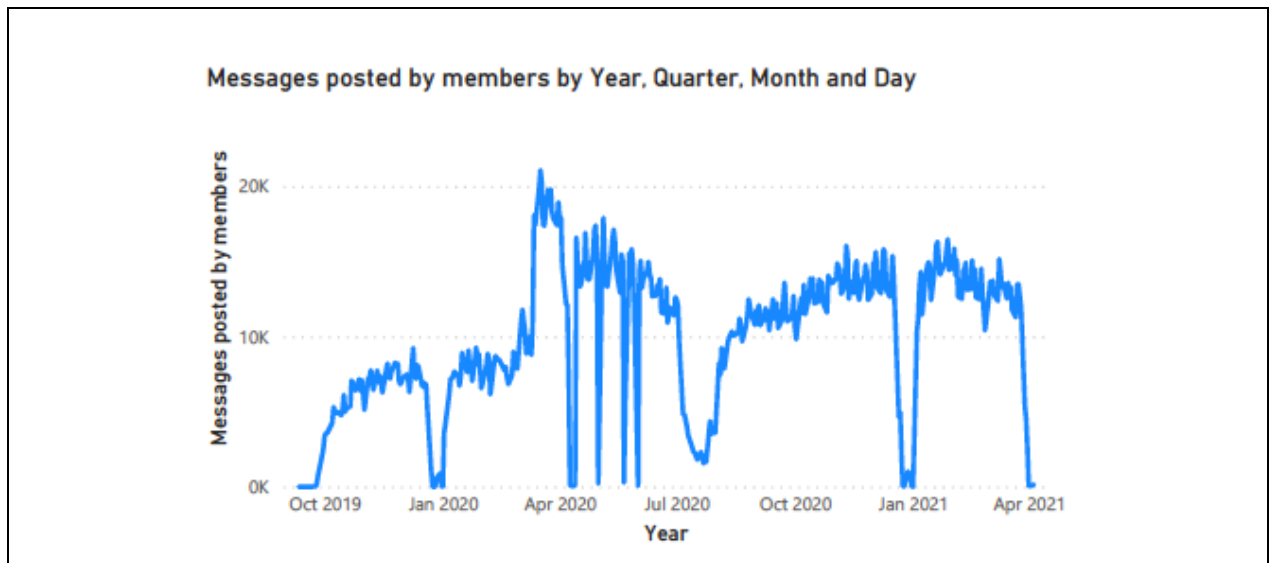

**Figure 3.  Direct messages posted by members**

### *Sociability – Teams Become More Task-oriented*

The data on sociability (Table 2) indicate that using the digital tools facilitated staying in touch with team members (item 1) and that people were quite comfortable with the digital work situation (item 8). Feelings of loneliness (item 2), ease of staying updated on team members (item 3) and the possibility of developing a well-functioning team (item 5) all received medium scores. The lowest scores were reported for having conversations that were not about work (item 9) and developing close friendships with team members in the digital work situation (item 10).

|  | N | Factor loading | Factor loading | Mean | Std. Dev. |
|---|---|---|---|---|---|
| 1. The digital work situation makes it easy for me to stay in touch with my team members. | 207 |  | .81 | 4.10 | 1.10 |
| 2. I do not feel lonely in this digital work situation. | 207 |  | .71 | 3.62 | 1.26 |
| 3. The digital work situation makes it possible for me to stay updated on my team members. | 207 |  | .74 | 3.52 | 1.11 |
| 4. The digital work situation enables spontaneous and informal conversations. | 207 | .59 |  | 3.01 | 1.30 |
| 5. The digital work situation enables us to develop a well-functioning team. | 207 | .56 | .54 | 3.43 | 1.11 |
| 6. The digital work situation enables me to develop good working relations with my team members. | 207 | .73 |  | 3.04 | 1.17 |
| 7. The digital work situation enables me to identify with the team. | 207 | .73 |  | 3.12 | 1.11 |
| 8. I am comfortable with the digital work situation. | 207 |  | .66 | 3.99 | 1.08 |
| 9. The digital work situation opens for conversation that is not about work. | 207 | .77 |  | 2.93 | 1.25 |
| 10. The digital work situation enables me to develop close friendships with my team members. | 207 | .82 |  | 2.23 | 1.20 |

**Table 2. Mean and standard deviations of sociability scores.**

We further conducted an exploratory factor analysis to assess the factor structure of the sociability scale. The results reveal a two-factor structure with factor loadings ranging from .56 to .82. The first factor consists of items 1–3, 5 and 8, which are directed at more task-oriented aspects, such as "The digital work situation makes it possible for me to stay updated on my team members." The second factor consists of items 4–7, 9 and 10, which are more relational-oriented, such as "The digital work situation opens for conversation that is not about work." Item 5, "The digital work situation enables us to develop a well-functioning team", cross-loaded on both factors with a factor loading of .56 on factor 1 (task-related) and .54 on factor 2 (relational). Considering that item 5 captures a general aspect of team functioning, it is reasonable that this item cross-loaded on both the task-related factor and the relational factor. To compare the two factors, we first removed item 5 from both and examined their mean difference. The mean value of task-related sociability with items 1–3 and 8 was 3.81 (std. dev. = 0.90), significantly higher than the mean value of relational sociability with items 4, 6–7 and 9–10 (mean = 2.87, std. dev. = 0.95) with a p-value less than .01. In other words, participants perceived that the digital tools were better at facilitating task-related sociability, such as staying in touch with the team, updates, etc., than facilitating non-work or relational sociability, such as spontaneous, informal, non-work-related communication. We proceeded further with our analyses using the two dimensions: task and relational sociability.

We conducted multilevel regression analyses to examine the use of electronic communication tools, such as emails, chats, telephone, video conferencing, shared documents and team coordination and their associations with the participants' perceived task-related and relational sociability. None of the digital communication tools was significantly related to task-related sociability. However, email was negatively related (-.19, p < .01), while chats were positively related (.30, p < .10) to relational sociability: the more team members relied on email as a communication channel to approach their work, the less they experienced social relations. However, the more they relied on chats, the more they could carry out informal and non-work conversations. Next, we looked at time spent on meetings. Neither hours spent on planned meetings nor hours spent on unplanned meetings were related to task or relational sociability.

Team tenure (i.e. how long a person had been on the team) did not have a direct relationship with task-related sociability (.00, $p > .10$) or relational sociability (.01, $p > .10$). However, combined with team coordination, team tenure helped explain how team members felt more relational sociability. Our multilevel analysis revealed a positive relationship between coordination and task-related sociability (.60, $p < .01$) and relational sociability (.37, $p > .10$). Furthermore, those with low coordination with their teams felt even less relationally sociable towards those with less tenure on their teams than those with more tenure. However, no moderating effect of team tenure was observed on the positive relationship between coordination and task-related sociability. We also observed variations between teams regarding their sociability scores; see Figures 4 and 5 below.

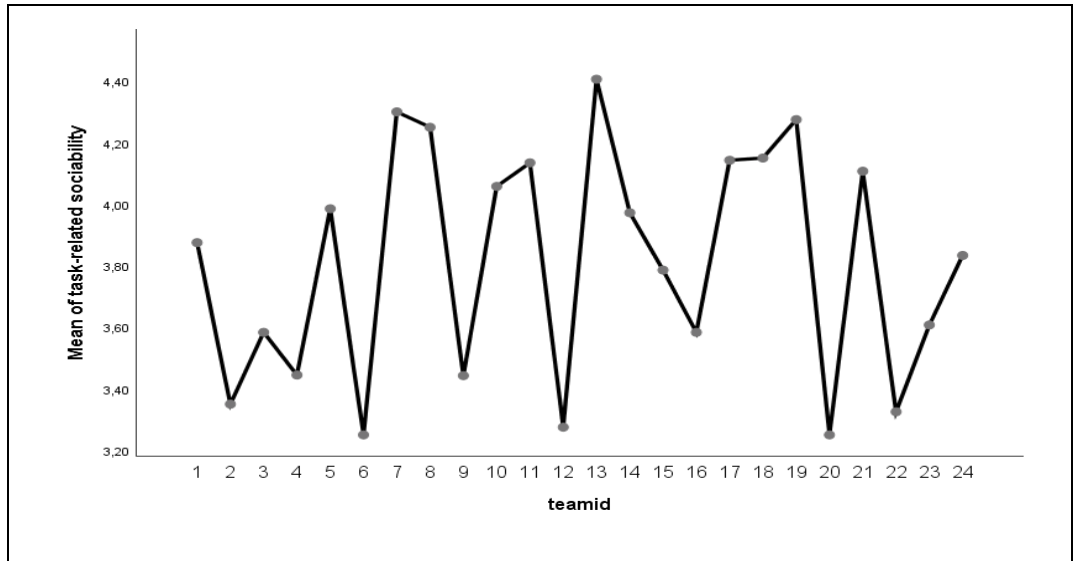

**Figure 4. Team variation on task-oriented sociability.**

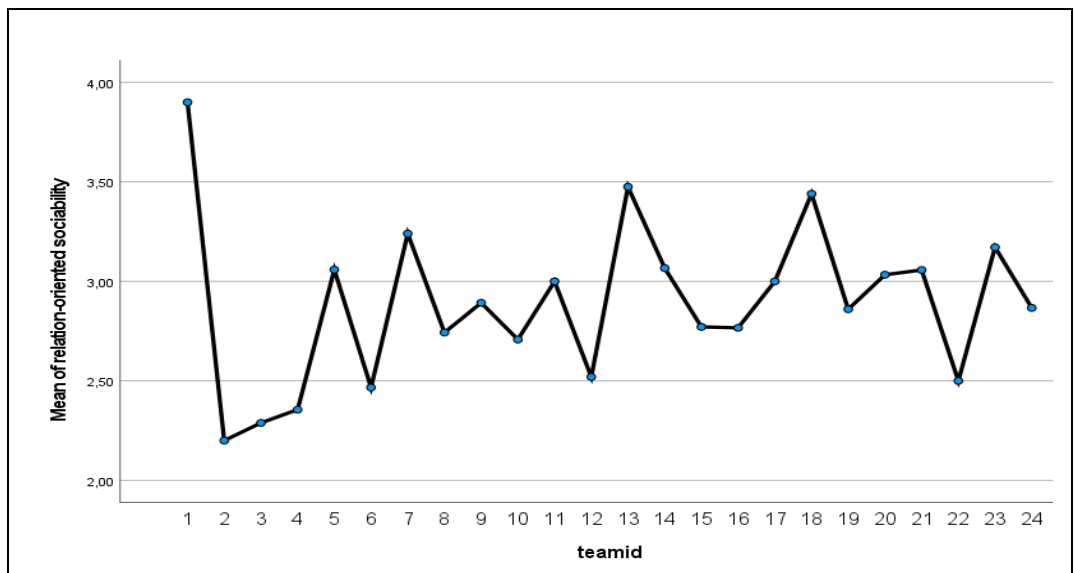

**Figure 5. Team variation on relation-oriented sociability.**

## *Variation: Personnel Changes and Team Organization*

We conducted interviews with Teams 2–7 and 10. Teams 2, 3, 4 and 6 had lower scores on the sociability scale, while Teams 5, 7 and 10 were higher. In the following sections, we present the findings from our interviews with the team leaders that indicate differences between the teams.

*Changes in personnel and responsibilities*

Overall, the teams with lower sociability scores (Teams 2, 3, 4 and 6) experienced more personnel changes before and during Covid-19, whereas the teams with higher scores were more stable in terms of personnel. Team 2, for example, was quite recently put together and started in December 2019, four months prior to the lockdown. Because of the Covid-19 lockdown, some team members had not even met other team members in person, something that was considered a challenge:

> "We have two team members that have not met others in the team in person. Even if you participate in the daily meetings, you perhaps do not get that team feeling". (Team leader 2)

On Team 6, most team members have been replaced since June 2019. Having quite a new team undermined the team members' senses of belonging when the team transitioned to distributed work, as the team leader explained:

> "Now, with even more people in the team, there will be a lack in the sense of belonging, quite simply. We haven't had time to settle in. I was trying to build a team, and then new members came in. There has been a lot of rotation [of members between teams] and onboarding, and I have spent a significant amount of time on that. There are only a few team members that have been here for many years". (Team leader 6)

Teams 4 and 6 also experienced changes in both personnel and responsibilities due to resource reallocation during Covid-19. Two software developers, one tester and one UX designer left Team 4, and the team's responsibilities shifted from developing new software to maintenance. This change in responsibilities was seen as less rewarding by the team members, as the team lead commented:

> "When it comes to coordinating the work and having clear goals, when the Corona situation came, we faced cut-downs in the budgets. We had clear goals and OKRs, but we no longer had the resources to reach [those] goals. These are the goals that we need to reach, not the other ones [i.e. they had to reprioritize their goals]. We have discussed whether it is something wrong with the goals or if rather we are not able to do what we are supposed to do". (Team leader 4)

Changes in personnel and responsibilities were also a concern for Team 6. Predominantly because of reduced budgets to cope with the financial impact of the Covid-19 lockdown, experienced consultants were replaced by new people with less experience. For instance, new testers came in who had to learn with what the team was working. Second, they had received an additional domain with which to work, but the number of team members remained the same, so team members had to take on extra tasks.

*Meetings and ceremonies*

The teams had many similar meetings and ceremonies, such as daily stand-ups, Monday commitments and Friday wins. Team 2 did not have one-on-one talks between the team leader and team members during the Covid-19 lockdown. Teams 2 and 3 have not had retrospectives during Covid-19, but all the other teams interviewed had retrospectives.

*Size and team organization*

There were also differences in the size and organization of the teams. The teams that scored higher on sociability were smaller or had sub-teams. For instance, Team 5 operated with three sub-teams.

Team 7 was smaller with six team members. There was a lot of individual work on this team, and the members described it as three 2-person sub-teams. The team leader claimed that they had not been concerned with working actively on onboarding:

> "I have been doing a poor job on that. Those who were at [Site 2] came with some "luggage". They came with a product that they had been working with […]. They could not focus on the team." (Team leader 7)

Team members had a high level of autonomy. There was no product owner, and the team members decided for themselves what to work on to a large degree. The team leader explained,

> "I have not been acting like a boss. I have let people decide for themselves what they think is interesting". (Team leader 7)

During this home office period, members of the team found it easier to work in pairs, as the team leader explained:

> "We have one-on-one meetings when people work ad hoc. When someone is stuck or needs someone to talk to, they call each other on Slack and share a screen. We simply work together in pairs and do pair programming. We have done pair programming to a larger degree now than before, I would say. You do not feel that you disturb anyone. Usually [when co-located], you would have walked over [to another team member], perhaps you would feel that you disturbed [that person], and you would have had to find a quiet room". (Team leader 7)

Team 10 was organized in sub-teams, such as areas of responsibility (e.g. login, native capabilities and native applications) and disciplines (e.g. iOS, Android, backend and security), and they worked on projects for the personal market, business market and effective development. The disciplines functioned as sub-teams, while members of disciplines worked dynamically across projects. The introduction of sub-teams to the larger team was considered beneficial:

> "It is a feeling of working more appropriately now, with disciplines and sub-teams. The first retrospectives I did after I joined [the unit] were about the need for smaller teams." (Team leader 10)

## Discussion and Implications

Research has demonstrated how agile ISD organizations create and share knowledge in practice (Orlikowski 2002) by face to face, tacit, and co-located work (Chau et al. 2003; Conboy 2009; Ghobadi and Mathiassen 2016). Such forms of knowledge creation and sharing are based on social ties and common ground. The Covid-19 pandemic changed this as people were, overnight, transitioning to distributed and digitally mediated work. While we know some about distributed ISD (Sarker and Sarker 2009) less is known about how agile ISD organizations cope when rapidly transitioning from co-located to distributed work. We therefore asked the research question - how do digital tools afford knowledge creation and sharing in agile ISD organizations in a sudden transition to distributed work? To answer this research question, we have reported findings from a large ISD organization delivering digital products to an alliance of banks. The significance of digital tools comes from our analytical lens of LSCs, as findings from LSCs indicate how the socio-technical affordance of digital tools matter for knowledge deliberation. Our aim is to focus on what is possible to achieve with distributed digital collaboration, rather than comparing it to the gold standard of face to face collaboration (Faraj et al. 2016). However, most LSC research does not address social ties and common ground, something which is central to knowledge creation and sharing in agile ISD organizations. We therefore also apply the perspective of sociability (Kreijns et al. 2007) in our analysis. Our findings indicate how: *i*) digital tools afford more task-orientation, and *ii*) different teams have different variations that need to be handled. We discuss how digital tools and practices must emerge together to maintain and create new sociability for agile ISD organizations.

*Digital tools afford increased task-orientation*

As a sudden transition such as the Covid-19 shutdown happens, its implications goes beyond a change in practice. The organization experiences severer changes. In this case a co-located, agile ISD organization takes on a form of LSC. This is because different forms of distance were introduced. Those with children at home doing home schooling found it difficult to work during regular office hours. This created a situation common to distributed ISD teams, where team members work in different time zones, influencing among others the amount of time they can work together (Sarker and Sarker 2009; Shameem et al. 2018). For teams used to being co-located, in addition to physical and temporal distance, even socio-cultural distances emerged. As an example, practitioners had to balance software development while simultaneously be a kindergarten teacher. It created a new form of role multiplexity (Maruping and Matook 2020). The significance of this lies in the observation that they went from a co-located situation where everyone was on the same page, also in terms of roles. Suddenly, practitioners no longer merely represented a team or the company, but their role in the cooperation changed as well. This resembles LSCs, where participants not only represent the collaboration at hand, but different interests and organizations, with different interests. The introduction of distance represented significant variation that had implication for knowledge creation and sharing.

Becoming a form of LSC, new challenges emerge for knowledge deliberation, such as reflective inquiry, abstract reasoning, debate, and dialogue on topics essential to the work being done (Malhotra et al. 2021). The relation between what the practitioners need to accomplish in terms of knowledge deliberations and the features of the digital tools becomes essential. A key issue for the organization under study during the immediate first period of the Covid-19 shutdown was to remain productive. This was necessary to keep the banks´ digital offerings up and running. There was a concern regarding how well they would be able to maintain software while being distributed, echoing research on distributed ISD (Zimmermann et al. 2018). Much new development was halted, however, some teams had to implement new digital offerings. An example of a new offering that was needed was a new loan solution that was to be provided by the government to support struggling business with emergency, government guaranteed, loans. There was a need for knowledge deliberations. Our findings show that the unit managed to maintain its productivity (see Figure 1). We also observe that knowledge deliberations moved to digital tools (see spike in March on Figures 2 and 3). With all the issues reported for distributed agile ISD, and the new challenges introduced in LSCs, it is reasonable to wonder why productivity did not decline. One reason for this could be that the teams in our study, although mostly co-located, were developing and maintaining software using the same digital tools before, during, and after the Covid-19 shutdown. They had a digital production toolchain using tools such as GitHub and Maven, and they used digital tools for deliberations, such as Confluence, Trello, and Slack. These are tools that typically afford what is needed in LSC knowledge deliberation. Such as identifying critical dependencies between people (e.g., who are notified if a code change breaks the build), and experimenting in parallel paths (Maven allows forking etc.). It is also reasonable to believe that they were reaping the benefits of already established social ties and the common ground established before the shutdown. It is relevant therefore to consider the changes in knowledge deliberations, what the tools afford to this end, and consider the wider implications with regard to maintaining and even creating new forms of sociability, as we do next.

Digital tools standardize knowledge deliberation in distributed ISD (Anwar et al. 2019; Balijepally and Nerur 2019). Couple this with the socio-technical observation that when speed or efficiency is required in an organization, practices and tools tend to become more standardized (Trist and Bamforth 1951). Our findings illustrate how the digital tools coupled with the need for productivity afforded a formalization of knowledge creation and sharing. They kept their knowledge deliberation practices, such as daily stand-up meetings, something important in knowledge deliberation in agile ISD (Chau et al. 2003, Conboy 2009, Ghobadi and Mathiassen 2016). Practices such as daily stand-ups, Monday commitments and Friday wins that used to be co-located were now done using digital tools like Slack and Teams. Our findings indicate that they had fewer such scheduled knowledge exchanges after the lockdown. The tools afford such scheduled knowledge exchanges. However, arranging scheduled online meetings is perhaps not the most challenging for agile ISD organizations. A more severe challenge is that we find that they previously had more informal knowledge exchanges (e.g., talks around the coffee machine, turning to the person at the nearby desk for a question, and chatting in the hallways going to meetings). Such informal deliberations now had to be scheduled on the calendar. The difference of which anyone who ever has attended an online session can attend to. They did several things to overcome this. They had social gatherings such as coffee breaks on Teams or arranged social quizzes using specific tools. We observe a separation between the formal knowledge deliberations (such as standups) and more social events (digital coffee breaks). They also used Slack for texting messages asking technical questions, but this also was considered a more formal form of communication and leave less room for relation-oriented communication. This is apparent in the Slack usage (Figure 2 and 3), where we find a predominance of private channels and direct messaging, effectively losing the informal knowledge exchanges that used to happen listening for example to chatter around the team areas (Smite et al. 2019). We consequently observe more of a separation between relation-oriented communication and task-oriented communication, as is expected in LSCs (Malhotra et al. 2021). The separation could mean situated and emergent knowledge deliberation, so important in agile ISD organizations suffers (Chau et al. 2003). In agile ISD, continuous knowledge creation happen on-action (e.g., daily stand-ups), however, equally important is the deliberations happening in-action, that is through solving practical problems together (Babb et al. 2014). As agile ISD organizations moves towards LSCs, one should be aware how sociability is maintained through practical problem solving using digital tools.

Investigating more in depth what the different tools afford in terms of sociability, the different types of digital tools used were not related to task-oriented sociability, but email was negatively associated with relational-oriented sociability, while chat (e.g., in Slack) was positively associated. In other words, reliance on email as a communication channel weakens social relations, whereas chat affords more informal and non-work communication. These tool affordances are particularly interesting considering the finding that neither hours spent on planned, nor unplanned meetings were related to task or relational sociability. This implies that the amount of time may not necessarily contribute to team sociability, but socio-technical practices do, including what communication behavior the digital tools afford (Krejins et al., 2007). For those agile ISD organizations concerned with keeping social ties and common ground, digital tools should support both kinds of sociability, so that there is a complementarity between a high task focus and a high social focus in communication. In traditional LSCs, such as open source communities, leading participants are recognized based on their expertise and knowledge, demonstrated by the quality of the code they contribute (Faraj et al. 2015). Leaders, allies, and knowledge collaborators are potentially identified differently as agile ISD organizations move towards LSCs, Agile ISD practitioners may need to maintain and create social ties for effective knowledge sharing (Balijepally and Nerur 2019; Kotlarsky and Oshri 2005). Maintaining and even developing sociability can be done employing prosocial behaviors as simple as thanking and including personal references in digitally mediated problem solving (Faraj et al. 2015).

As knowledge deliberations changed, teams and practitioners were encouraged to experiment with new ways of creating and sharing knowledge using digital tools. It is therefore relevant to see what practitioners reported digital tools afforded in this new setting. Data from the sociability questionnaire indicate that team members perceived the digital tools to afford task-related sociability, such as staying in touch with the team, updates, etc., more than facilitating non-work or relational sociability, such as spontaneous, informal, non-work-related communication. While digital tools do allow for relational communication, doing so constitutes a new socio-technical practice. Before the shutdown, the digital tools were primarily used for task-related sociability, and all the relational-oriented sociability happened face to face in the office. Incorporating relational-related sociability into the digital tools is a new practice that needs time and effort to be established and domesticated (Faraj et al. 2016).

*Digital tools must support a wide range of variations*

In our study, teams varied in size, tenure, practices, and management approaches. During the Covid-19 lockdown, the teams were encouraged to keep their practices but also to experiment actively with new practices and adapt to the situation. The teams were allowed to experiment with interaction patterns that created social spaces for team members to overcome the challenges of being distributed (Kotlarsky & Oshri, 2005).

As our data show, teams scored differently on the sociability scale, but not just because of different configurations of digital tools. The multiple regression analysis results show that a lower degree of coordination leads to a lower degree of relational sociability for newer team members. This is consistent with our observations in the interviews, which indicated that the teams that reported relatively lower on sociability were either new teams or had new team members. Team 2, for example, was quite new, and some team members had not even met any of their team members in person. It is likely that for new team personnel, it is more difficult to communicate about social aspects using digital tools (Slack, for instance).

Our findings indicate that team size and organization may also influence sociability. The teams with relatively higher sociability scores (5, 7 and 10) were small teams or teams organized into smaller sub-teams. Team 7, for example, had six team members, and their informal way of organizing with high individual autonomy seems to have transitioned well to distributed work. They experienced cooperation as easier digitally, as they could work in pairs enabled by the shared screen function and did not have to find a quiet room, as they used to when co-located. While we lack data that can explain these differences, we can speculate that smaller groups, particularly if they are used to working together, face less of the knowledge deliberation challenges that larger teams face when transitioning to an LSC context. They can freeride, as it were, on the established social ties and common ground. Larger teams and teams with new members have a closer resemblance to LSCs and are more likely to face the knowledge deliberation challenges associated with LSCs, such as separating productive task conflict from interpersonal conflict, and lack of shared memory (Malhotra et al. 2021). However, as larger teams must focus on using tools to achieve relational sociability, smaller teams should focus on maintaining their ability for knowledge deliberations over time as variations emerge. As variances occur in all work systems (Trist and Bamforth 1951), agile ISD organizations moving towards LSCs should have an increased awareness of how their practices and digital tools work to either reduce variation (i.e. increase speed and efficiency) or increase variation (i.e. nurture diversity and stimulating new ideas).

*Practical implications*

Our findings have some practical implications for agile ISD organizations moving towards LSCs. First, teams that are used to working with a digital toolchain (e.g., Maven, GitHub, etc.) and collaborating digitally (e.g., Slack) seem able to maintain task-oriented communication. However, even for teams that are digitally proficient, this is no guarantee that they are competent in the relational aspects of digital communication. Teams that are fluent in using digital tools for production should focus on incorporating sociability into their practical problem solving.

Second, appreciate that each team is different and allow socio-technical practices to emerge. There is probably no one-size-fits-all approach to dealing with radical changes, such as the Covid-19 lockdown. Teams need guidance, and goals, but they should have freedom to experiment with practices and tools that fit their context, tasks, and responsibilities. The case reported here suggests that, in times of radical change, regular team health checks are important to allow team members to express how they experience the work situation. If there is a challenge, teams follow up with structured problem solving and continuous improvement. There are also measurements of how much software is produced. Teams and management can keep track of goals and accomplishments using OKRs and practices such as Monday commitments and Friday wins. Authority should be granted to teams to find ways to maintain sociability and knowledge creation and sharing based on social ties.

Third, a crisis such as the Covid-19 lockdown does not only force distributed work; other changes occur as well, such as changes of personnel, tasks, and responsibilities. Follow-up and reflection in one-on-one conversations and reflections (such as retrospectives) at the team level are necessary to avoid team members feeling left alone with the challenges imposed by these changes.

*Limitations*

Our empirical work has limitations. It is based on a single case study and company and criticism concerned with uniqueness of findings apply to our study. While we do not aim to make generalizable statements, the large number of teams investigated in the company, and the use of several data sources lessen validity concerns.

# Conclusion and Future Work

The results of this study indicate that the knowledge deliberation challenges agile ISD teams face when transitioning to distributed set-ups are not necessarily related to task-oriented practices, but rather the relational aspects of agile ISD. To confront these challenges, there probably will be no one-size-fits-all approach; it depends on the variation teams face, such as tenure, size, digital fluency, and knowledge deliberation needs such as speed versus creativity. Managers in agile organizations are recommended to cultivate the relation-oriented communication where practitioners can build social ties while solving practical problems. Maintaining and establishing new social ties will become a pertinent challenge for agile ISD organizations also in the coming period of hybrid work (partially at the office, partially at home). As such, new venues for future research opens. We have only begun to consider the implications of agile ISD organizations taking on forms of LSCs, and the implications for knowledge creation and sharing. More studies are needed to empirically detail the implications and provide research-based advice. Beginning with the observation that is possible to achieve social ties in digitally supported LSCs, studies can provide insight into how this can be done. How do agile ISD organizations as LSCs emerge differently from open-source communities or outsourcing arrangements? Also, what are the best, digitally mediated practices for knowledge creation and sharing? As dedicated digital social arrangements can have limited effects, how can one bring a social orientation into the actual problem-solving activities? Answering such questions seem valuable for a future of hybrid agile ISD.

# Acknowledgement

This research is funded by the Research Council of Norway through the 10xTeams project (grant 309344)